\documentclass[twocolumn,showpacs,preprintnumbers,amsmath,amssymb,pre,nofootinbib]{revtex4-1}

\usepackage[utf8]{inputenc}
\usepackage[T1]{fontenc}
\usepackage{multirow}
\usepackage{mathptmx}
\usepackage[export]{adjustbox}
\usepackage{mathtools}
\usepackage{float}
\usepackage{color}    
\usepackage{graphicx}
\usepackage{dcolumn}
\usepackage{bm}
\usepackage{subfigure}
\usepackage{amssymb}
\usepackage{amsmath}
\graphicspath{{plots/}}

\usepackage{siunitx}
\DeclarePairedDelimiterXPP\BigOSI[2]%
  {\mathcal{O}}{(}{)}{}%
  {\SI{#1}{#2}}
  
\begin{document}

\title{ Ion core effect on transport characteristics in warm dense matter}

\author{T.S. Ramazanov}
\affiliation{ Institute for Experimental and Theoretical Physics, Al-Farabi Kazakh National University, 71 Al-Farabi ave., 050040 Almaty, Kazakhstan}

\author{M. Issanova}
\affiliation{ Institute for Experimental and Theoretical Physics, Al-Farabi Kazakh National University, 71 Al-Farabi ave., 050040 Almaty, Kazakhstan}

\author{Ye.K. Aldakul}
\affiliation{ Institute for Experimental and Theoretical Physics, Al-Farabi Kazakh National University, 71 Al-Farabi ave., 050040 Almaty, Kazakhstan}

\author{S.K. Kodanova}
 \email{kodanova@physics.kz}
\affiliation{ Institute for Experimental and Theoretical Physics, Al-Farabi Kazakh National University, 71 Al-Farabi ave., 050040 Almaty, Kazakhstan}


\begin{abstract}
An effective potential approach in combination with the molecular dynamics (MD) method was used to study the effect of the ionic core on 
the transport properties of ions in the warm dense matter regime.
As an example, we considered shocked silicon.
The  results of MD simulations within microcanonical ensemble were analyzed by computing the mean squared displacement (MSD) and the velocity autocorrelation
function (VAF) of particles. The MSD and VAF are used to compute the diffusion coefficient of ions. 
The results are compared with the data computed neglecting the ion core effect.
It is found that the ion core effect leads to a significant decrease of the diffusion coefficient.  
Additionally, we computed the viscosity coefficient of ions using the Green-Kubo relation connecting viscosity and the stress autocorrelation function.
It is revealed that the ion core effect can cause increase or reduction of the viscosity coefficient  depending on the strength of inter-ionic coupling.

\end{abstract}

\maketitle

\section{\label{Sec. I} Introduction}

An interest in warm dense matter (WDM), a state with high temperature and density, is fueled due to development of new facilities where extreme conditions are created by laser driven heating and shock compression (e.g. see Refs. \cite{PhysRevE.81.036403, doi:10.1063/1.5143225} for the discussion of relevant system parameters). These experiments are motivated by the fact that WDM is found in various astrophysical objects, such as giant planets \cite{Saumon, Militzer, Guillot}, white and brown dwarfs \cite{Saumon,Becker}, and interiors of stars \cite{Daligault}. 
Additionally, warm dense plasma is generated in experiments on inertial thermonuclear fusion \cite{Hu}. 
An example of experimental discovery that have been made in this field is diamond formation due to irradiation of material surfaces with intense plasma flows, etc. \cite{Kraus1,Kraus2}. 
However, a comprehensive understanding of fundamental physical properties of WDM is often hindered by complexity of the required simulations \cite{burke-thermal-dft_13, Moldabekov_SciPost_2022, doi:10.1063/1.5097885, doi:10.1063/1.5003910}. Particularly, the study of the transport properties of ions is challenging due to required long simulation times, e.g. compared to the calculation of structural properties.
One possible way to overcome this difficulty is to use an effective potential, extracted from ab initio density functional theory based simulations \cite{Vorberger2012, doi:10.1063/5.0059297, PhysRevE.81.065401, PhysRevE.103.L051203, PhysRevB.103.165102} or derived using density response function of electrons \cite{https://doi.org/10.1002/ctpp.202000176, doi:10.1063/1.4829042, issanova_2016, https://doi.org/10.1002/ctpp.201500137, https://doi.org/10.1002/ctpp.201400094, https://doi.org/10.1002/ctpp.201700109, doi:10.1063/1.4922908, https://doi.org/10.1002/ctpp.201700113, doi:10.1063/1.4932051},   in  molecular dynamics \cite{PhysRevE.99.053203, Vorberger2012}.

The computation of transport properties of WDM and dense plasmas is of relevance for practical applications such as inertial confinement fusion \cite{Kritcher, Hurricane}.
Thus, various approached were used to compute different transport coefficients in WDM regime, e.g. see the recent review by Grabowski et al \cite{ Grabowski_HEDP}.
To further deepen our understanding of the physics of transport phenomena at WDM parameters, in this work we perform  study of the viscosity and diffusion coefficients in the WDM of shocked silicon.
Moreover, for the first time, to our knowledge, we provide an analysis of  a transition from ballistic-type diffusion on short time scales to normal diffusion on long time scales taking into account the ion core effect.
To capture relevant physical effects, we use the effective potential from Ref. \cite{Vorberger2012}, which provides adequate description of structural and dynamical properties of shocked silicon at WDM conditions. The effective potential is used to model ions by the molecular dynamics (MD) method in the microcanonical ensemble.
The latter point is important since it allows us to avoid possible artificial effects due to nonphysical forces in often used schemes with thermostat such as Langevin dynamics and Nosé-Hoover dynamics.

Following Ref. \cite{Vorberger2012},  the effective ion-ion potential used in this work reads
\begin{equation}
\label{eq01}
V(r)=\frac{Z_i^2 e^2}{4 \pi \epsilon_0 r} \exp \left(-k_s r\right)+\frac{\left(Z_n^2-Z_i^2\right) e^2}{4 \pi \epsilon_0 r} \exp \left[-\left(b+k_s\right) r\right],
\end{equation}
here  $Z_i$ denotes the ion charge state, $k_s$ is the inverse screening length, $Z_n$ is  the nuclear charge, and parameter $b$ determines the range of the short-range repulsion.
From Eq. (\ref{eq01}) one can see that the effective ion-ion interaction is represented by Yukawa type potential with additional short-range repulsion (Yukawa+SRR).

In Eq.(\ref{eq01}), the first term is often referred as the Yukawa potential and takes into account the screening of ion-ion interaction in the long-wavelength limit.
The second term in Eq.(\ref{eq01}) accounts for the bound core electrons and represents what we call the ion core effect in this work.

The aim of this work  is the investigation of the ion core effect on the self diffusion process and viscosity at characteristic WDM parameters.
To study the diffusion process and compute diffusion coefficient we analyzed the mean squared  displacement and velocity auto-correlation function of ions.
The latter is used to compute diffusion coefficient using Einstein’s relation. The correctness of the value of the diffusion coefficient computed this way is validated by comparing to the data for  the diffusion coefficient extracted from the mean squared  displacement of ions on long time scales.
The viscosity coefficient is computed using the Green-Kubo relation connecting viscosity coefficient and the microscopic stress tensor.

The paper is organized as the following:
In the next section \ref{sec: parameters} we provide information about used simulation parameters.
Then in Sec. \ref{Sec. II}, the results of the MD simulations based on the effective potential (\ref{eq01}) are presented.
The paper is concluded by summarizing our main findings.

\section{\label{sec: parameters} Simulation parameters}

We took the system parameters as in Ref. \cite{Vorberger2012}, i.e. electron density $n_e=5.36 \times 10^{23}~\text{cm}^{-3}$ and temperature $T_e= 54540 ~{\rm K}$, $Z_n=14$, $Z_i=4$, and mean inter-particle distance $a=(3/4 \pi n_i)^{1/3}=2.291~a_0$, where $a_0$ is the Bohr radius and $n_i$ is the ion density.
We set $b=0.7~a_0^{-1}$ and $k_s=1.277~a_0^{-1}$.
From latter we obtain the value for screening parameter $\kappa=k_sa=2.926$.
The system temperature is controlled by coupling parameter $\Gamma=Z_i^2e^2/4 \pi \epsilon_0 a k_B T_i$, where $T_i$ is the temperature of ions.
At $T_i=T_e$, we have $\Gamma=40.4$.
Since WDM state with different temperatures of ions and electrons (i.e. non-isothermal state) is often generated in experiments \cite{PhysRevE.98.023207}, where ions can be hotter or colder than electrons depending on the way of generation of WDM, we additionally considered different $\Gamma$ values
from $1$ to $100$. These correspond to ion temperatures in the range between $1.9~{\rm eV}$ and $188~{\rm eV}$. This serves our objective which is the elucidation of the ion core effect on transport properties in different ionic correlation regimes.


The investigation was conducted by means of MD simulations.
The dynamics of $N=103823$ identical particles was obtained by solving the equation of motion through Beeman's algorithm with a time step of $t = 0.01\omega_p^{-1}$.
We employed periodic boundary conditions with a main box of side length $L=(4 \pi N/3)^{1/3} a$.
The results are presented in dimensionless units; dimensions of length are reduced by $a$ and a time unit is represented by the inverse plasma frequency of ions $\omega_p^{-1}$.
The results from MD simulations are measured within the microcanonical ensemble. This allows us to avoid possible nonphysical effects due to the use of thermostat which, for example, is represented by an additional artificial friction force and compensating it randomly fluctuating force in the case of the Langevin dynamics. 
We note that such methods like the Langevin dynamics and the Nosé-Hoover dynamics allow one to generate correct ionic configurations within the canonical ensemble, but the transport and dynamic properties can be incorrect since they dependent on actual evaluation of particles trajectories in time.

\section{\label{Sec. II} MD simulation results}

\subsection{Radial distribution function}

\begin{figure}[b]
\includegraphics[width=\linewidth]{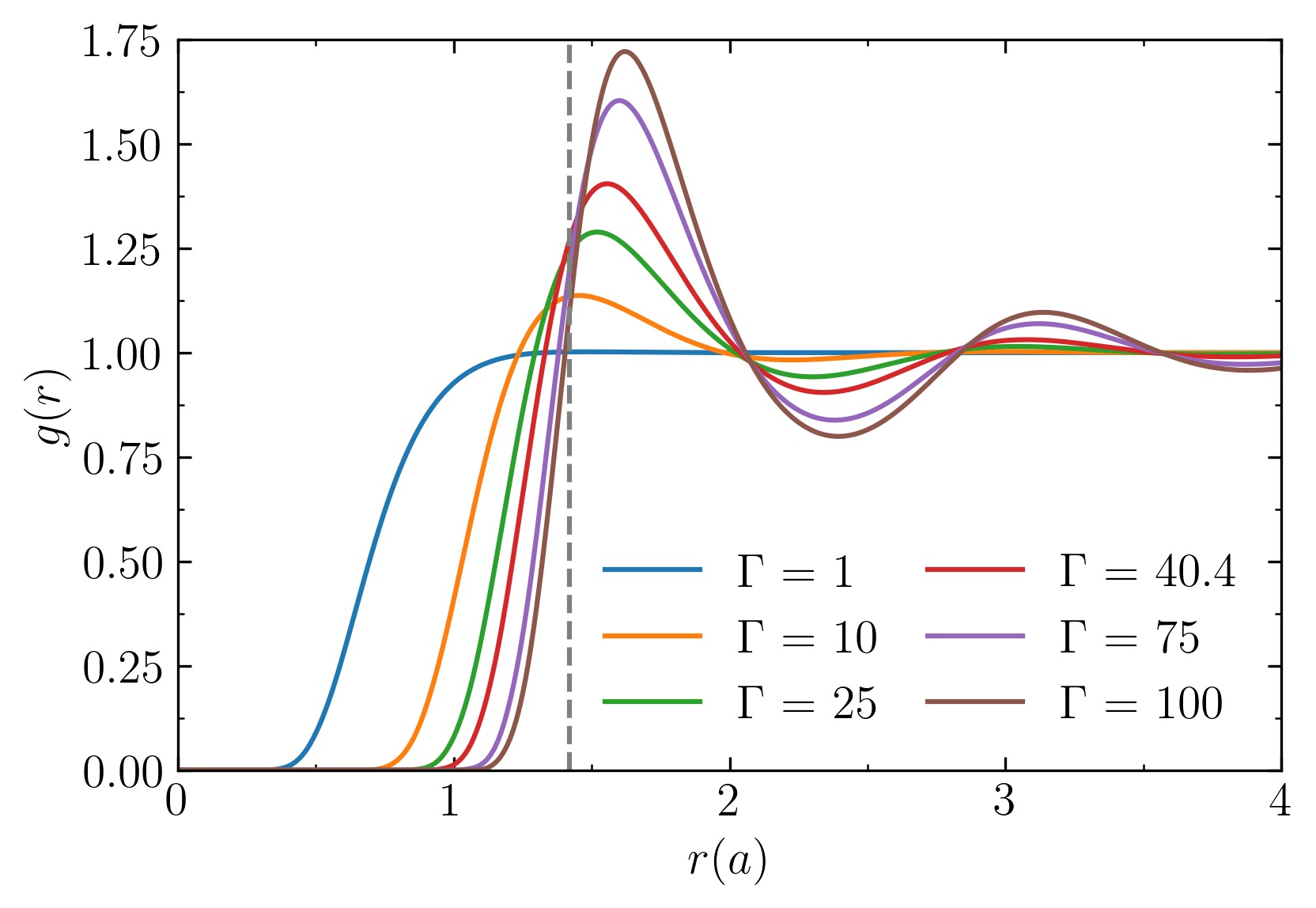}
\caption{ Radial distribution function  of ions computed using Eq. ~(\ref{eq01}) at different coupling parameters.}
\label{fig01}
\end{figure}

To have better understanding of the results obtained for various considered properties, we calculated the radial distribution function (RDF).
The RDFs for a range of coupling parameters are shown in Fig. \ref{fig01}.
From Fig. \ref{fig01} we see that at $\Gamma=1$ the system shows no significant order in the structure besides of a correlation hole.
With increasing value of $\Gamma$ the RDF shows the emergence of a short-range order, which is indicated by the distinct peaks in the RDF.
In Fig. \ref{fig01}, the vertical dashed line at $r/a=1.42$ represents a characteristic range of the short-range repulsion defined by $b^{-1}$.
At all considered $\Gamma$ values, there is a significant probability of positioning ions at a distance within range  $b^{-1}$.
Thus, we can expect that the ion core effect is important for transport characteristics at all considered $\Gamma$ values.

\subsection{Diffusion characteristics}

To analyze the effect of an additional repulsion due to ionic core on the diffusion coefficient, we computed  the mean squared displacement (MSD) of ions  and the velocity autocorrelation function (VACF) of ions.
To cover both short- and long-time correlations with a desired accuracy at a minimum computational effort we implemented the modification of the order-$n$ algorithm \cite{Dubbeldam2009} for computing correlation functions, originally proposed by Frenkel and Smit \cite{Frenkel2002}.
The gain in computational effort is possible due to the use of a variable sampling frequency, i.e. short- and long-time correlations are calculated at different sampling frequencies.
Implementing a conventional method with the fixed-frequency sampling \cite{Frenkel2002,Rapaport2004} is impractical as the number of particles gets larger as the desired correlation times become longer.

In most cases the MSD and VACF of a single MD run contain a large statistical noise.
To reduce the undesired statistical noise one usually runs several independent MD simulations under the same set of parameters and take the average of the results.
Thus, to reduce the statistical noise, the results of the MSD and VACF presented in this paper have been averaged over five MD runs.

We note that the MSD and VACF contain information about microscopic dynamics of particles. 
The MSD represents how a particle is displaced [in a random direction] from a given position at a given arbitrary moment in time.  In general, the MSD contains more information than the diffusion coefficient, see e.g. Ref. \cite{msd_cpp_09}. Information about the diffusion coefficient is extracted from the MSD by analyzing its behavior at long times, which is relevant, e.g., for hydrodynamics.
Similarly, in addition to the diffusion coefficient, the VACF contains information about wavenumber-averaged oscillation (excitation) modes \cite{PhysRevE.56.7310}.

The MSD is calculated using the following equation:
\begin{equation}
\label{eq02}
u(t)=\left\langle|\mathbf{r}(t)-\mathbf{r}(0)|^{2}\right\rangle,
\end{equation}
where $\langle\dots\rangle$ denotes ensemble average.

\begin{figure}
\includegraphics[width=\linewidth]{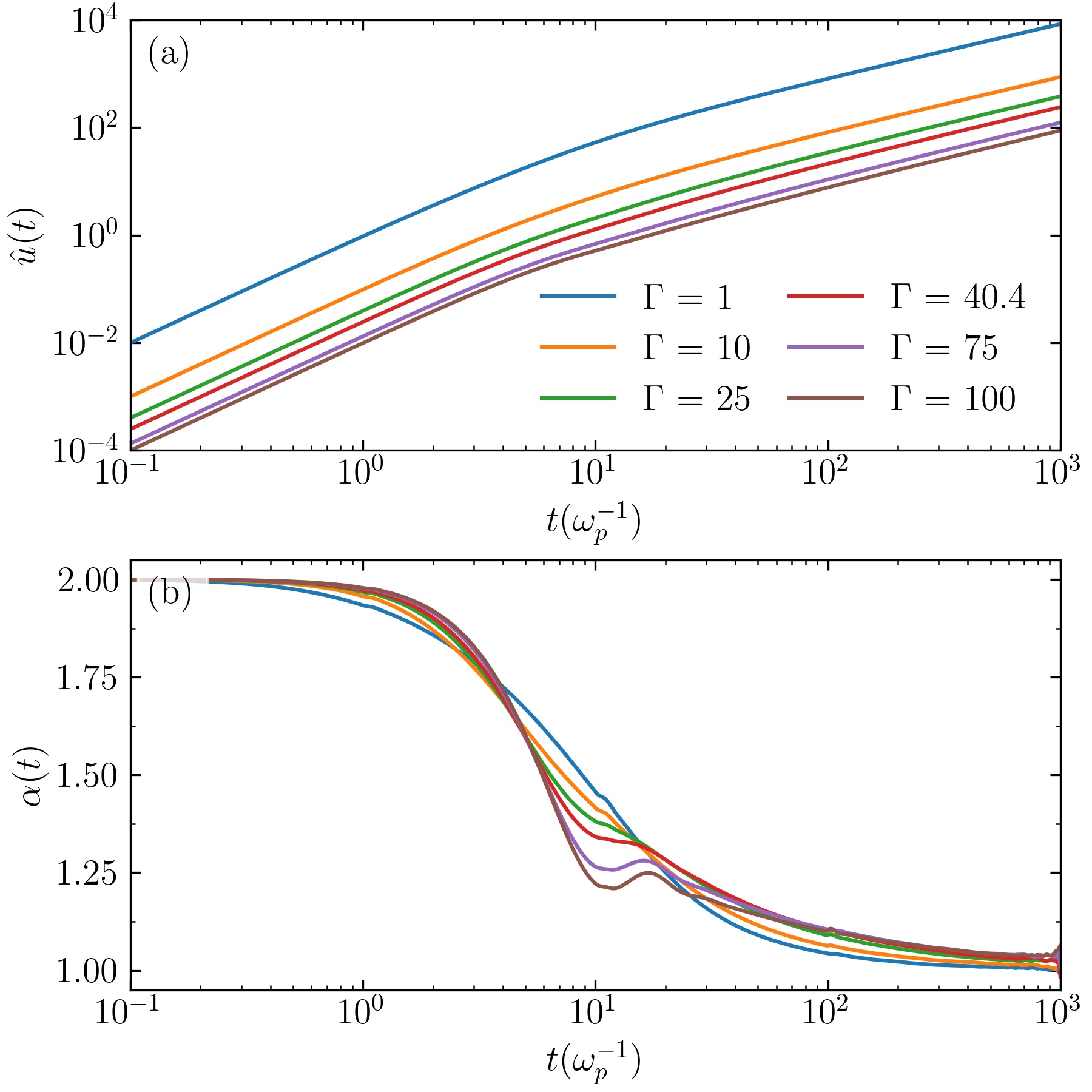}
\caption{(a) Reduced Mean Squared Displacement (MSD) of ions with the bound core electrons at different coupling parameter values.
(b) Diffusion parameter $\alpha (t)$ of a system of ions as defined by Eq. (\ref{eq03}), where $\alpha$ is dimensionless quantity.}
\label{fig02}
\end{figure}

The results of simulations for different values of $\Gamma$ are presented in Fig. \ref{fig02}. 
From  Fig. \ref{fig02} we see that,  as one would expect, the emergence of order in the system due to the increase of the coupling parameter causes the MSD of particles to decrease.
Indeed, at larger $\Gamma$ values, ions become more confined by neighboring particles and the displacement process becomes less intensive.

In the case of normal diffusion, the MSD must be a linear function of time $u(t)\propto t$. From  Fig. \ref{fig02} (a) one can see that it is the case at long times $t> 100 \omega_p^{-1}$. At shorter times the deviation from normal diffusion takes place. For an accurate calculation of the diffusion coefficient, the MSD values should be used at times scales where the normal diffusion regime is well established. 
For this propose, the deviation from normal diffusion can be  analyzed by considering an effective time dependent diffusion parameter $\alpha$ as it is defined through the proportionality
\begin{equation}
\label{eq03}
u(t)\propto t^{\alpha}.
\end{equation}

\begin{figure}
\includegraphics[width=\linewidth]{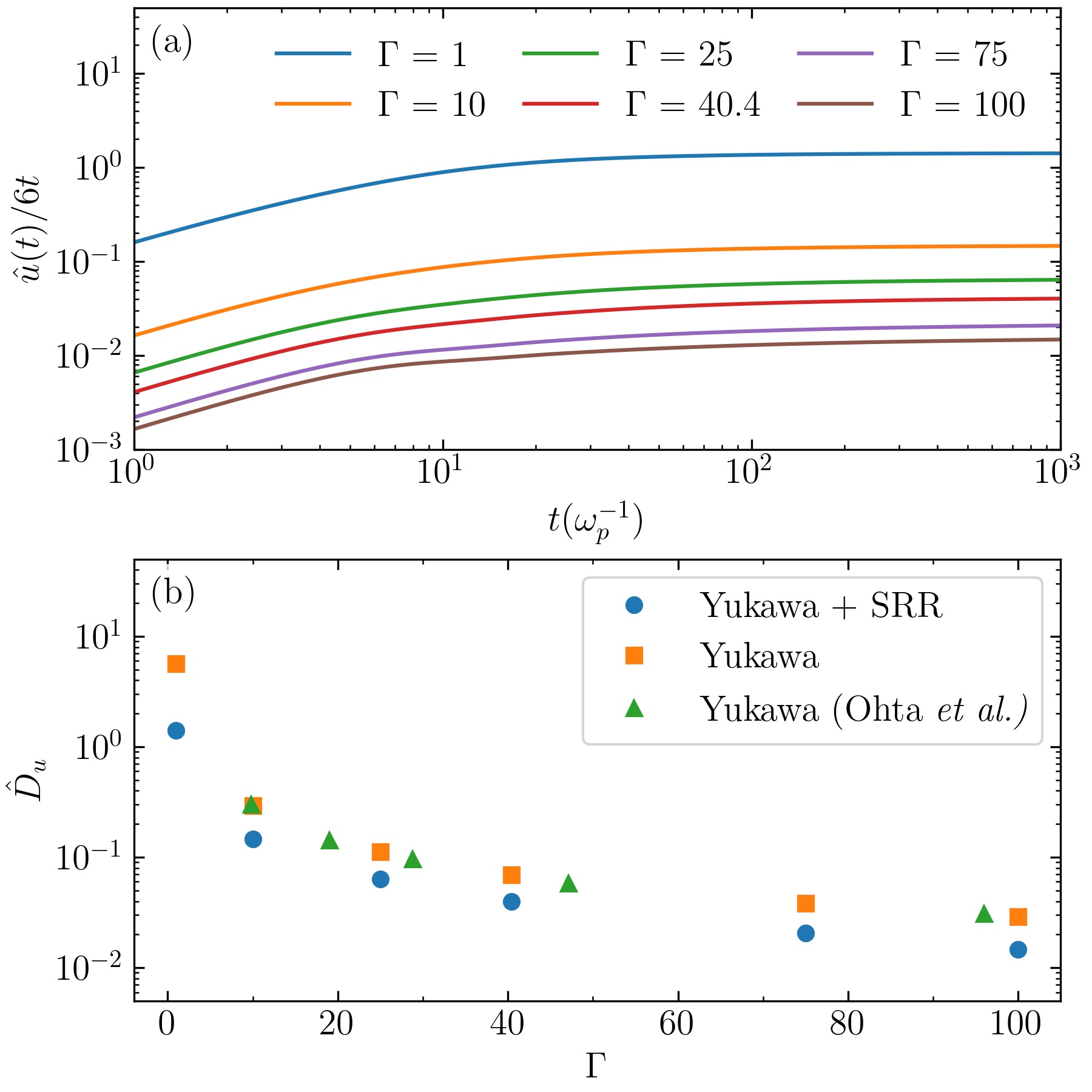}
\caption{(a) The ratio $\hat{u}(t)/6t$ as a function of time which shows convergence to a diffusion coefficient, where $u(t)$ is the MSD in units of the square of the mean interparticle distance, $a^2$, and $t$ is in units of the inverse ion plasma frequency. (b) The diffusion coefficient values (in units of $\omega_p a^2$) calculated by the Einstein’s relation using the Yukawa+SRR potential, Eq. (\ref{eq01}), and the standard Yukawa potential.
The comparison with the results for Yukawa systems reported by Ohta et al. \cite{Ohta2000} is also shown.}
\label{fig03}
\end{figure}

\begin{figure}
\includegraphics[width=\linewidth]{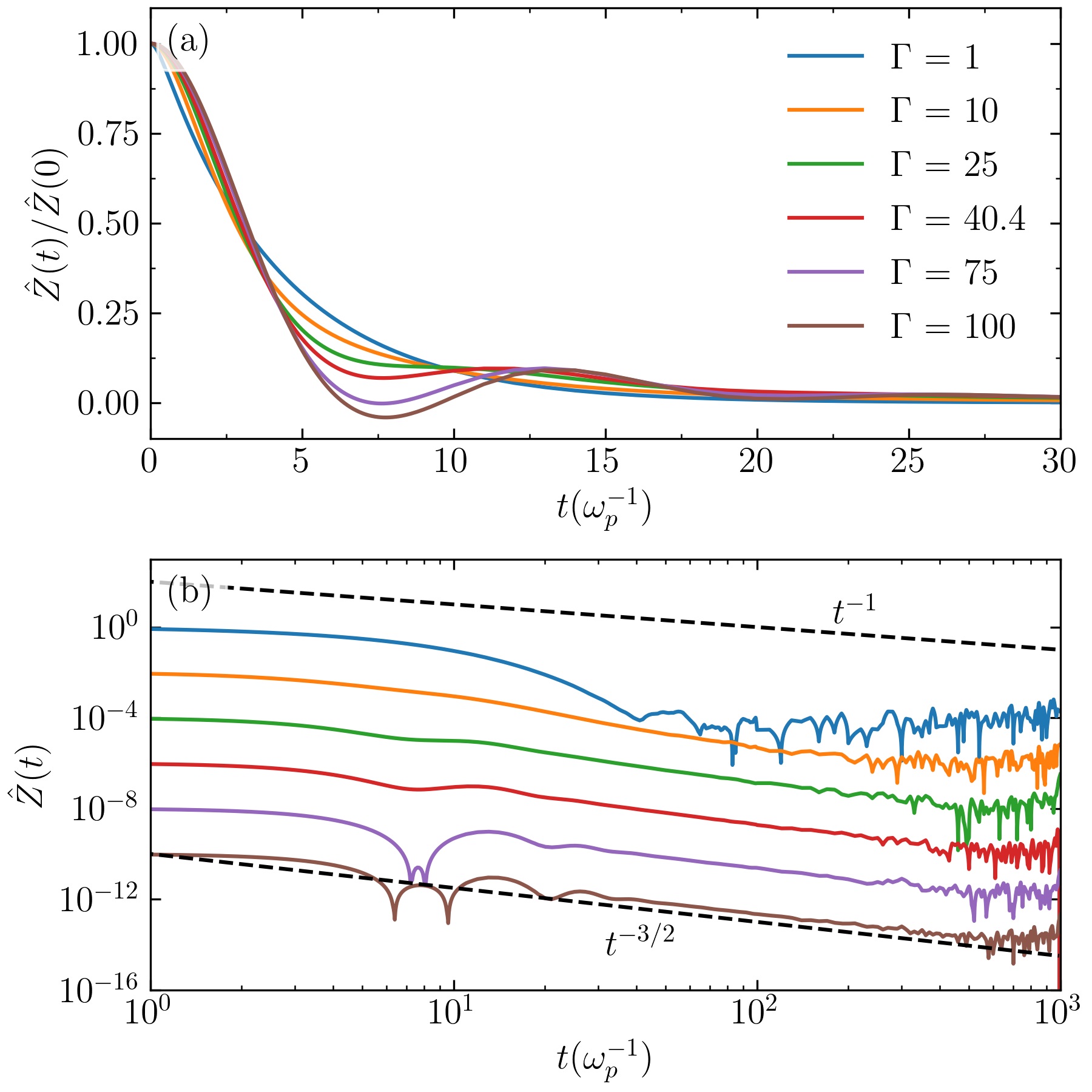}
\caption{(a) Velocity autocorrelation function (VACF) of ions at different values of the  coupling parameter.
(b) The reduced VACF in logarithmic scale where dashed line represents $t^{-1}$ decay. In the subplot (b), the curves of the VACF at $\Gamma>1$ are shifted down for better visualisation of the behavior at long times.}
\label{fig04}
\end{figure}

\begin{figure}
\includegraphics[width=\linewidth]{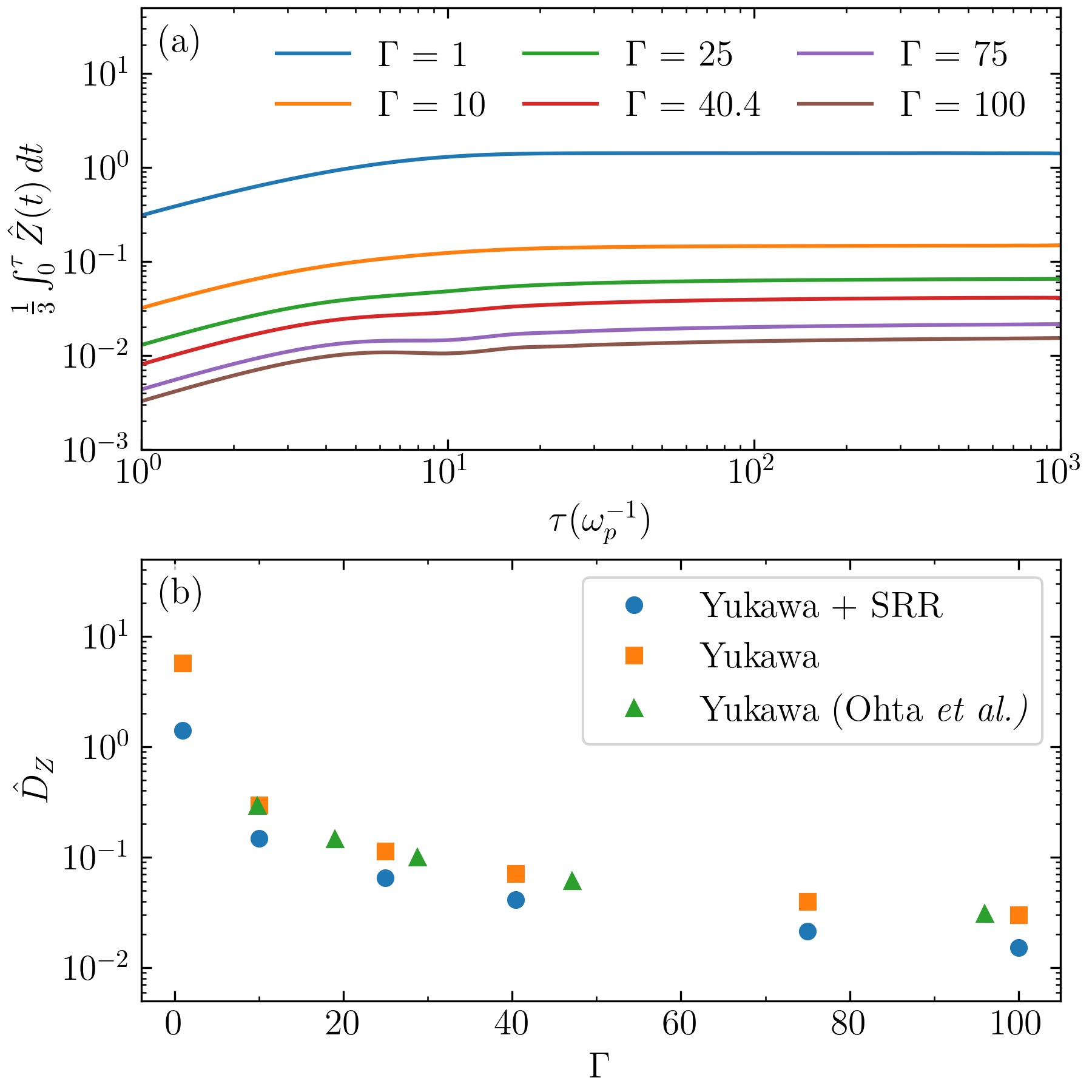}
\caption{(a) The VACF integral as a function of time, which shows
convergence as the upper limit of integration increases. (b) The diffusion coefficient values  (in units of $\omega_p a^2$) calculated by using the Green-Kubo relation
 on the basis of  the Yukawa+SRR potential, Eq. (\ref{eq01}), and the standard Yukawa potential. The comparison with the results
for Yukawa systems reported by Ohta et al. \cite{Ohta2000} is also shown.}
\label{fig05}
\end{figure}

Fig. \ref{fig02} (b) shows the diffusion parameter obtained from the analysis of the MSD data presented in Fig. \ref{fig02} (a).
The plot of the diffusion parameter shows that the initial stage of the motion is ballistic, $\alpha=2$, from which it switches to the intermediate so-called anomalous diffusion, $\alpha\ne1$, and then converges to the normal diffusion, $\alpha=1$.
We also observe that an intermediate diffusion type and the rate of convergence to the normal diffusion are controlled by the value of coupling parameter, the increase in the coupling parameter hinders the convergence to a normal diffusion. Clearly, for capturing a normal diffusion  regime in the case of strongly coupled ions characteristic for WDM, one needs to perform measurements at times scales $t\gg 100~\omega_p^{-1}$, where $\omega_p^{-1}$ is the plasma period of ions. We note that this represents a significant challenge for a standard Kohn-Sham density functional theory based MD simulations, where computation time  of electronic structure at each MD step at considered high temperatures represents a major bottleneck. This is an example where alternative simulation techniques of ions dynamics such as effective potential approach or orbital free density functional method are indispensable.

Next, at long time scales, the values of the diffusion coefficients can be obtained using Einstein's relation:
\begin{equation}
\label{eq04}
D_u=\lim_{t\to\infty}\frac{u(t)}{6t}.
\end{equation}

Fig. \ref{fig03} (a) shows the ratio $\hat{u}(t)/6t$. From Fig. \ref{fig03} (a) we see that $\hat{u}(t)/6t$ approaches to a constant value at $t\gg 10^{2}~\omega_p^{-1}$.
However, as it can bee seen from  Fig. \ref{fig02} (b),  we could not  reach  exactly normal diffusion regime with $\alpha \equiv 1$ for $\Gamma>1$.
Thus, there is certain error in the diffusion coefficient values computed using Eq. (\ref{eq04}).  
Using Eq. (\ref{eq04}) and the average of $u(t)/6t$ in the time interval $500<t\omega_p<1000$, we have calculated the approximate values of the diffusion coefficients summarized in Table \ref{table01}.

\begin{table*}
\caption{Diffusion coefficients of a system of ions with interaction potential (\ref{eq01}) and pure Yukawa potential.
Einstein's and Green-Kubo relations are used to calculate the diffusion coefficients from MSD and VACF data, respectively.
Error bounds correspond to 95\% confidence interval.}
\label{table01}
\begin{ruledtabular}
\begin{tabular}{ccccc}
    $\Gamma$ & $D_u/\omega_p a^2$ & $D^{\mathrm{Yukawa}}_u/\omega_p a^2$ & $D_Z/\omega_p a^2$ & $D^{\mathrm{Yukawa}}_Z/\omega_p a^2$\\ \hline
    $ 1 $ & $ 1.409 \pm 0.003 $ & $ 5.62 \pm 0.06 $ & $ 1.412 \pm 0.005 $ & $ 5.7577 \pm 0.007 $ \\
    $ 10 $ & $ 0.1455 \pm 0.0008 $ & $ 0.294 \pm 0.002 $ & $ 0.1475 \pm 0.0007 $ & $ 0.2973 \pm 0.001 $ \\
    $ 25 $ & $ 0.0633 \pm 0.0006 $ & $ 0.1121 \pm 0.0008 $ & $ 0.0649 \pm 0.0003 $ & $ 0.1142 \pm 0.0002 $ \\
    $ 40.4 $ & $ 0.0399 \pm 0.0005 $ & $ 0.0697 \pm 0.0007 $ & $ 0.0411 \pm 0.0001 $ & $ 0.0713 \pm 0.0002 $ \\
    $ 75 $ & $ 0.0206 \pm 0.0003 $ & $ 0.0385 \pm 0.0005 $ & $ 0.0214 \pm 0.0002 $ & $ 0.0399 \pm 0.0002 $ \\
    $ 100 $ & $ 0.0146 \pm 0.0002 $ & $ 0.0289 \pm 0.0005 $ & $ 0.0152 \pm 0.0002 $ & $ 0.0302 \pm 0.0003 $ \\
\end{tabular}
\end{ruledtabular}
\end{table*}

The decrease of the diffusion coefficient values with increasing coupling parameter at $\Gamma>1$ is caused by the emerging local order and caging of an ion by  surrounding particles.
From Table \ref{table01} we observe that the ion core effect leads to significantly lower values of the diffusion coefficient. This can be understood as the result of stronger inter-particle correlations and stronger caging of an ion by its surrounding.

As mentioned, an error in the calculation of the diffusion coefficient is caused by the deviation of the MSD from the linear dependence on time.
In order to cross check the accuracy of our calculations, we used the VACF and the related Green-Kubo equation to computed the diffusion coefficient of ions.

The VACF is a measure of the correlation degree in the velocity of a particle at different times and is defined as:
\begin{equation}
\label{eq05}
Z(t)=\left\langle\mathbf{v}(t)\cdot\mathbf{v}(0)\right\rangle.
\end{equation}

The results for the VACF of ions at different values of the coupling parameter are shown in the top panel of Fig. \ref{fig04}.
At $\Gamma=1$, the VACF decays monotonically with time. This behavior changes with the increase in $\Gamma$ and we observe appearance of oscillatory pattern. 
The diffusion coefficient can be calculated by integrating the VACF over time. This is known as a Green-Kubo relation, which reads:
\begin{equation}
\label{eq06}
D_Z=\frac{1}{3}\int_{0}^{\infty}Z(t)dt,
\end{equation}
where we introduced notation $D_Z$ in order to distinguish it from the diffusion coefficient $D_u$ computed using the MSD. 

To find accurate data for the diffusion coefficient, the integral in Eq. (\ref{eq06}) must converge well.
In the bottom panel of Fig. \ref{fig04} we show the VACF using logarithmic scale to better illustrate the behavior of the VACF at long times.
At all considered $\Gamma$ values, the tail of the VACF decays faster that $t^{-1}$ which ensures the convergence of the integral in Eq. (\ref{eq06}).
However, one can observe from Fig. \ref{fig04} (b) that the increase in $\Gamma$ leads to a slower decay of the VACF values at long times.
For example, at $\Gamma=100$, the VACF decays as $t^{-1.5}$ at long times as it is illustrated in Fig. \ref{fig04} (b).
At $\Gamma \geq 10$, one need to generate data at times $t \gg 10^2 \omega_p $. 
As time increases, significant data fluctuations appear at some point due to a finite number of particles in the MD simulations.
Thus, for the calculation of the diffusion coefficient,  it is important that these fluctuations begin after the VACF value is sufficiently reduced so that they do not affect the integration accuracy in Eq. (\ref{eq06}). 

Due to the use of large number of ions in the main cell in our simulations, the convergence of the integral $\int_0^{\tau}Z(t)dt$ is not affected by fluctuations as it is demonstrated in Fig. \ref{fig03} (c).
The results for the diffusion coefficient $D_Z$ computed using  (\ref{eq06}) are summarized in Table \ref{table01}, where we again compare with the results computed using the Yukawa potential. From this figure we see the same behavior of the diffusion  coefficient $D_Z$ as that of $D_u$.
Table \ref{table01} summarizes the diffusion coefficients calculated by Einstein's and Green-Kubo relations. 
From this table we see that computed $D_Z$ and $D_u$ values  are in agreement within evaluated statistical uncertainty.
The disagreement between $D_Z$ and $D_u$ is about $1\%$ at $\Gamma=1$ and $\Gamma=10$, about $2\%$ at $\Gamma=25$ and $\Gamma=40.4$, and about $4\%$ at $\Gamma=75$ and $\Gamma=100$.

\subsection{Simulation results for viscosity }

For the calculation of the viscosity coefficient, we use the Green-Kubo relation connecting viscosity and stress autocorrelation function (SAF):
\begin{equation}
\label{eq07}
\eta=\frac{1}{k_{B} T V} \int_{0}^{\infty} H(t) d t,
\end{equation}
where in order to improve statistics, for the SAF we use:
\begin{equation}
    H(t)=(H^{xy}+H^{yz}+H^{zx})/3,
\end{equation}
as the $x$, $y$, and $z$ directions are equivalent \cite{Saigo2002}.
 
For example, $zx$ component of the SAF is defined as
\begin{equation}
H^{zx}(t)=\langle\sigma^{z x}(t) \sigma^{z x}(0)\rangle,
\end{equation}
with $\sigma^{z x}$ being the $zx$ component of the microscopic stress tensor:
\begin{equation}
\sigma^{z x}(t)=\sum_{i=1}^{N}\left[m v_{i, z} v_{i, x}-\frac{1}{2} \sum_{j=1 \neq i}^{N}\frac{z_{i j} x_{i j}}{r_{i j}} V^{\prime}\left(r_{i j}\right)\right].
\end{equation}
The $xy$ and $yz$ components of the SAF and corresponding  microscopic stress tensors are defined in a similar way.

During MD simulations, we computed the dynamics of 103823 particles after equilibration for a total time of $10^3~\omega_p^{-1}$. Further, the presented results are averaged over five MD runs.

The results of the computations are presented in Fig. \ref{fig04}. 
The Fig. \ref{fig04} (a) shows the SAF for different coupling parameters.
From this figure we observe that one needs a high quality data for the SAF at least up to $10 \omega_p^{-1}$ for the evaluation of the viscosity using Eq. (\ref{eq07}). 
At considered simulation parameters, nonphysical fluctuations in the data for the SAF become dominant at $t\gtrsim 20 \omega_p^{-1}$.
However, these fluctuations do not lead to a significant deterioration of the accuracy of the integration in Eq. (\ref{eq07}).
This can be seen in Fig. \ref{fig04} (b), where the values of the integral $\int_{0}^{t} H(\tau) d \tau$ are shown for different values of the upper integration limit $t$.
From Fig. \ref{fig04} (b) we see that the integral $\int_{0}^{t} H(\tau) d \tau$ is converged at $t\simeq 20 \omega_p^{-1}$ and further increase of the upper integration limit $t$ leads to certain 
oscillations aground  a mean value due to aforementioned  pure statistics at long times. A maximum value of the  amplitude  of these oscillations is used to evaluate  uncertainty in our data for the viscosity coefficient.  

\begin{figure}
\includegraphics[width=\linewidth]{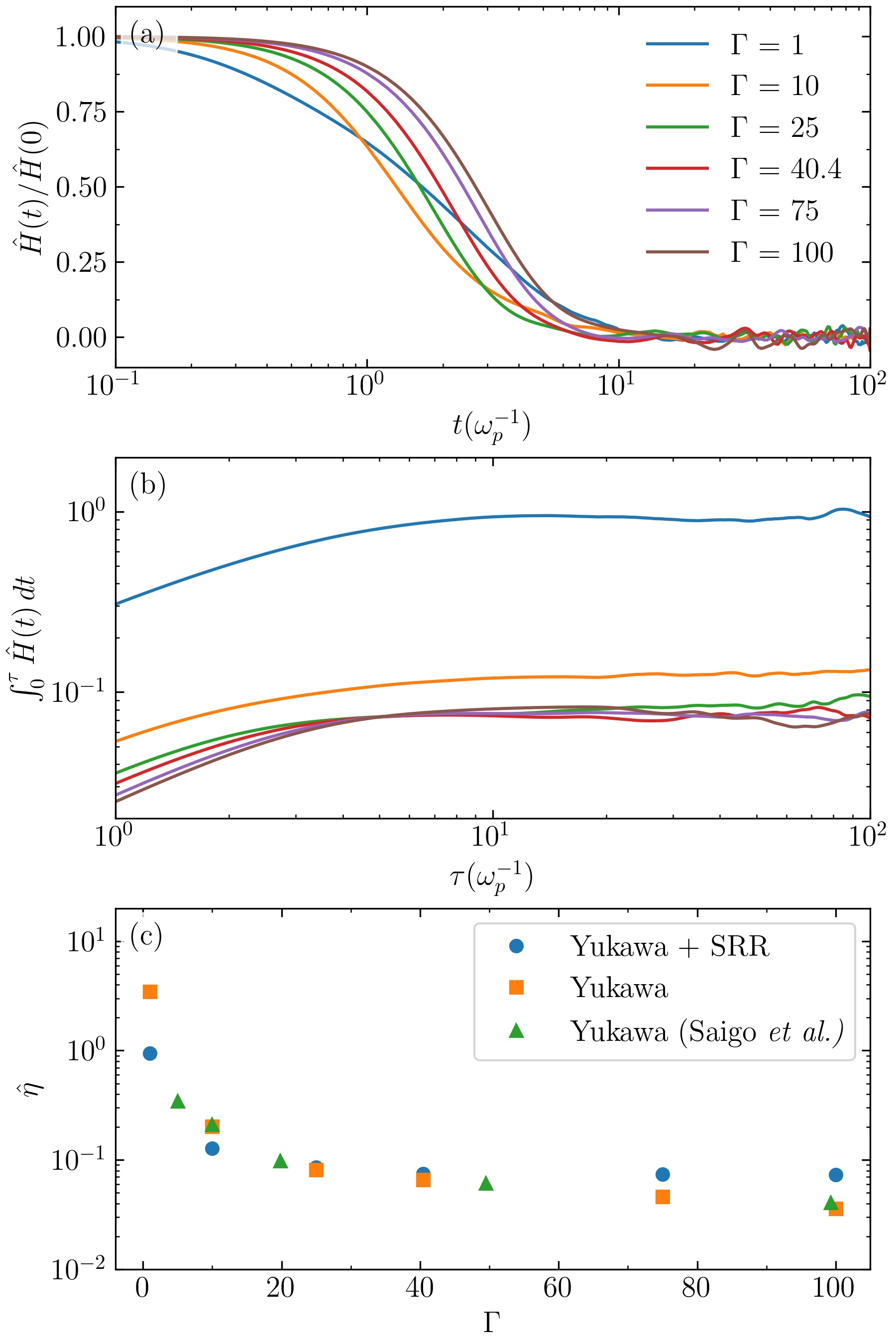}
\caption{(a) The  stress autocorrelation function (SACF) of ions at different values of the coupling parameter. (b) The integral of the reduced SACF as a function of time, which shows convergence as the upper limit of integration increases. (c) The viscosity values (in units of $n_i m_i\omega_p a^2$ with $m_i$ being the ion mass) calculated by using
the Green-Kubo relation , Eq. (7), on the basis of  the Yukawa+SRR potential, Eq. (\ref{eq01}), and the standard Yukawa potential. The data is compared with the results for
Yukawa systems reported by Saigo et al. \cite{Saigo2002}. }
\label{fig06}
\end{figure}

The results for the viscosity coefficient are summarized in Table \ref{table02} and compared with the data obtained for the Yukawa system.
At considered values of the coupling parameter, the viscosity decreases with an increase in $\Gamma$.
From the comparison with the data computed using the Yukawa potential, we see that at relatively small values of the coupling parameter $\Gamma<25$, the ion core effect leads to a decrease in the viscosity value. In contrast, at large values of the coupling parameter, the ion core effect leads to an increase in the viscosity value.
This can be understood considering different terms in Eq. (\ref{eq07}). 
As it is known \cite{PhysRevE.78.026408}, at relatively small values of $\Gamma$, the contribution coming from the term $m v_{i, z} v_{i, x}$ in the summation  in Eq. (\ref{eq07}) (a kinetic part) is dominant over the contribution due to term $\sum {z_{i j} x_{i j}}/{r_{i j}} V^{\prime}\left(r_{i j}\right)$ (a correlation part).
In contrast, at large $\Gamma$ values, the correlation part is dominant over the kinetic part in Eq. (\ref{eq07}).
At $\Gamma<25$, a stronger ion-ion correlation due to the ion core effect leads to a reduction of the mobility of ions and, thus, reduction of the the kinetic part of  Eq. (\ref{eq07}). As the result,
the ion core effect leads to a decrease in the viscosity value. At $\Gamma>25$,  the ion core effect induced stronger ion-ion correlation results in an increase in the correlation part in Eq. (\ref{eq07}). Thus, at large values of $\Gamma$, the ion core effect causes  an increase in the viscosity value.

\begin{table}
\caption{Viscosity of a system of ions with interaction potential (\ref{eq01}) and pure Yukawa potential. The Green-Kubo relation was used to calculate the viscosity from SACF data.
Error bounds correspond 95\% confidence interval.}
\label{table02}
\begin{ruledtabular}
\begin{tabular}{ccc}
    $\Gamma$ & $\eta/n_i m_i \omega_p a^2$ & $\eta^{\mathrm{Yukawa}}/n_i m_i \omega_p a^2$\\ \hline
    $ 1 $ & $ 0.94 \pm 0.09 $ & $ 3.5 \pm 0.3 $ \\
    $ 10 $ & $ 0.127 \pm 0.005 $ & $ 0.24 \pm 0.05 $ \\
    $ 25 $ & $ 0.087 \pm 0.01 $ & $ 0.081 \pm 0.006 $ \\
    $ 40.4 $ & $ 0.076 \pm 0.007 $ & $ 0.066 \pm 0.007 $ \\
    $ 75 $ & $ 0.074 \pm 0.004 $ & $ 0.046 \pm 0.004 $ \\
    $ 100 $ & $ 0.071 \pm 0.01 $ & $ 0.0358 \pm 0.006 $ \\
\end{tabular}
\end{ruledtabular}
\end{table}

The presented data Table \ref{table02} have uncertainty in the range from about $5\%$ up to about $14\%$ depending on the coupling parameter.
Although it is less accurate than the data obtained for the diffusion coefficient in this work, the uncertainty of the computed values of the viscosity still allows us to clearly distinguish the change in the general trend caused by the ion core effect when compared to the results obtained using the Yukawa potential.

\section{Conclusion}

In this work we have analyzed the  effect of the ion core  on the diffusion and viscosity of ions in the WDM.
We used the effective ion-ion interaction potential designed to describe the dynamic and static properties of shocked silicon. 
For the calculation of the diffusion coefficient we used two different approaches. First of all, we used the MSD of ions to compute the diffusion coefficient at long time scales.
Additionally, we used the VACF of ions to find the diffusion coefficient employing  the Green-Kubo relation connecting the VACF with the diffusion coefficient. 
These two methods allowed us to crosscheck our results for the diffusion coefficient.  
It was found that the ion core effect leads to a significant decrease of the diffusion coefficient at $1 < \Gamma\leq 100$.
This is in agreement with previously reported observations, e.g. from  simulations using an orbital-free density functional theory at the Thomas-Fermi-Dirac level \cite{Ticknor}.  
Similar analysis of the viscosity coefficient computed using corresponding Green-Kubo relation shows that the ion core effect leads to  the reduction of viscosity at $\Gamma<25$ and to the increase of the viscosity at $25<\Gamma\leq 100$. 

The performed analysis of the MD simulations results contributes to  our understanding of the transport properties of the WDM.
Moreover, it demonstrates the need of large scale MD simulations for an accurate calculation of transport properties of non-ideal ions in the WDM regime and, thus, highlights the need of fast MD simulation methods for further theoretical analysis of the WDM. 

\section*{Acknowledgments}
This research is funded by the Science Committee of the Ministry of Education and Science of the Republic of Kazakhstan Grant AP08856650 "Study of the structural, transport, and thermodynamic properties of non-ideal multicomponent dense plasma with heavy ions".

\bibliography{Main}
 
\end{document}